\documentclass[12pt,letter]{article}
\usepackage{graphicx, epsfig, color}
\usepackage{amssymb,amsmath}

\def\eslt{E_T^{\rm miss}}
\def\delew{\Delta_{\rm EW}}

\def\delhs{\Delta_{\rm HS}}
\def\delbg{\Delta_{\rm BG}}
\def\to{\rightarrow}

\def\bi{\begin{itemize}}
\def\ei{\end{itemize}}

\def\th{\tilde h}

\def\tst{\tilde t}

\def\tg{\tilde g}

\def\tq{\tilde q}
\def\tw{\widetilde W}
\def\tz{\widetilde Z}
\def\alt{\lesssim}
\def\agt{\gtrsim}
\def\be{\begin{equation}}  
\def\ee{\end{equation}}  
\def\bea{\begin{eqnarray}}  
\def\eea{\end{eqnarray}}  
\def\beas{\begin{eqnarray*}}  
\def\eeas{\end{eqnarray*}}
\def\mllcut{m_{\ell\ell}^{\rm cut}}
\newcommand\prd[3]{{\it Phys.\ Rev.\ }{\bf D #1} (#2) #3}

\newcommand\prl[3]{{\it Phys.\ Rev.\ Lett.\ }{\bf #1} (#2) #3}
\newcommand\plb[3]{{\it Phys.\ Lett.\ }{\bf B #1} (#2) #3}
\newcommand\jhep[3]{{\it J. High Energy Phys.\ }{\bf #1} (#2) #3}
\newcommand\app[3]{{\it Astropart.\ Phys.\ }{\bf #1} (#2) #3}

\newcommand\npb[3]{{\it Nucl.\ Phys.\ }{\bf B #1} (#2) #3}
\newcommand\epjc[3]{{\it Eur.\ Phys.\ J. }{\bf C #1} (#2) #3}

\newcommand{\hepph}[1]{hep-ph/#1}

\begin{document}
\begin{titlepage}
\begin{flushright}
FTPI-MINN-14/28\\
UH-511-1237-14
\end{flushright}

\vspace{0.5cm}
\begin{center}
{\Large \bf  Monojet plus soft dilepton signal\\
from light higgsino pair production at LHC14
}\\ 
\vspace{1.2cm} \renewcommand{\thefootnote}{\fnsymbol{footnote}}
{\large Howard Baer$^{1,2}$\footnote[1]{Email: baer@nhn.ou.edu },
Azar Mustafayev$^3$\footnote[2]{Email: azar@phys.hawaii.edu } and
Xerxes Tata$^3$\footnote[4]{Email: tata@phys.hawaii.edu } 
}\\ 
\vspace{1.2cm} \renewcommand{\thefootnote}{\arabic{footnote}}
{\it 
$^1$Dept. of Physics and Astronomy,
University of Oklahoma, Norman, OK 73019, USA \\
}
{\it 
$^2$William I. Fine Theoretical Physics Institute, 
University of Minnesota, Minneapolis MN 55455, USA \\
}
{\it 
$^3$Dept. of Physics and Astronomy,
University of Hawaii, Honolulu, HI 96822, USA \\
}
\end{center}

\vspace{0.5cm}
\begin{abstract}
\noindent 
Naturalness arguments imply the existence of higgsinos lighter than
200-300~GeV. However, because these higgsinos are nearly mass
degenerate, they release very little visible energy in their decays, and
(even putting aside triggering issues) signals from electroweak higgsino
pair production typically remain buried under Standard Model
backgrounds.  Prospects for detecting higgsino pair production via
events with monojets or mono-photons from initial state radiation are
also bleak because of signal-to-background rates typically at the 1\% level. 
Here, we consider the possibility of reducing backgrounds by
requiring the presence of soft daughter leptons from higgsino decays in
monojet events.  We find that LHC14 experiments with an integrated
luminosity of 300~fb$^{-1}$ should be sensitive to light higgsinos at
the 5$\sigma$ level for $\mu < 170$~GeV with a $S/B \sim 8.5$\%. For an
integrated luminosity of 1000~fb$^{-1}$ (which should be possible at a
high luminosity LHC) the corresponding sensitivity to $\mu$ extends to
over 200~GeV though the systematic uncertainty would have to be
controlled to considerably better than 5\%. The corresponding reach
from measurements of the rate asymmetry between monojet
events with same flavour vs. opposite flavour dileptons 
is $\sim$10-15 GeV smaller, but does not suffer the systematic
uncertainty from the normalization of the background.

\vspace*{0.8cm}

\end{abstract}

\end{titlepage}

\section{Introduction}

The lack of super-partner signals from new physics searches at 
LHC8 \cite{atlas_susy,cms_susy} (the LHC with $\sqrt{s}=8$~TeV), 
together with the measured value of the mass of a Standard Model (SM)-like 
Higgs boson \cite{atlas_h,cms_h}, has led
many authors to question whether weak scale supersymmetry (SUSY) as
realized by the Minimal Supersymmetric Standard Model is natural: 
{\it i.e.} whether the weak scale exemplified by the values of $M_Z$ or $m_h$
can be obtained without large fine-tuning of model parameters. 
The relative insensitivity of weak scale physics to physics at very high
scales (such as {\it e.g.}  the unification scale $M_{\rm GUT}$ in SUSY
Grand Unified Theories) was for many years (and remains) one of the
driving motivations for weak scale SUSY. We stress that weak scale SUSY
theories do not suffer from the huge fine-tuning problem of the Standard
Model (SM): while fine-tuning in the SM is a part in $10^{26}$,
fine-tuning in weak scale SUSY theories, by any of several measures
discussed in the literature, is typically no more than a part in ${\cal O}(10^4)$, 
and often significantly smaller~\cite{bbm,bbmp}.

The well-known relation \cite{wss},
\be 
\frac{M_Z^2}{2} = \frac{m_{H_d}^2 + \Sigma_d^d - 
(m_{H_u}^2+\Sigma_u^u)\tan^2\beta}{\tan^2\beta -1} -\mu^2 
\label{eq:mz}
\ee 
obtained from the minimization of the (renormalization group improved)
one-loop electroweak scalar potential of the Minimal Supersymmetric
Standard Model (MSSM) enables us to define $\delew$, which measures the
degree of cancellation between various contributions (defined at the
weak scale) to obtain the measured value of $M_Z^2$. In (\ref{eq:mz}),
$\Sigma_u^u$ and $\Sigma_d^d$ are radiative corrections which depend
strongly on the value of third generation squark masses.  Expressions
for the $\Sigma_u^u$ and $\Sigma_d^d$ are given in the Appendix of Ref.~\cite{rns}.  
The electroweak fine-tuning
parameter $\delew$ is defined by \cite{ltr,rns,perelstein},
\be 
\delew \equiv max_i \left|C_i\right|/(M_Z^2/2)\;, 
\ee 
where $C_{H_d}=m_{H_d}^2/(\tan^2\beta -1)$,
$C_{H_u}=-m_{H_u}^2\tan^2\beta /(\tan^2\beta -1)$ and $C_\mu =-\mu^2$.
Also, $C_{\Sigma_u^u(k)} =-\Sigma_u^u(k)\tan^2\beta /(\tan^2\beta -1)$
and $C_{\Sigma_d^d(k)}=\Sigma_d^d(k)/(\tan^2\beta -1)$, where $k$ labels
the various loop contributions included in Eq.~(\ref{eq:mz}).
Requiring weak scale
fine-tuning smaller than $\delew^{-1}$ then implies that,
\be
|\mu|^2 < \delew \frac{M_Z^2}{2}\;,
\label{eq:bound}
\ee
implying that higgsinos cannot be very heavy~\cite{ccn}.\footnote{We have tacitly
  assumed that higgsinos obtain their mass from the superpotential $\mu$
  term that then also enters the scalar Higgs potential. This is the
  case in all models we know of.}

We emphasize that the existence of light higgsinos, $\tz_{1,2}$ and
$\tw_1^\pm$, is a very robust feature of SUSY models with low
fine-tuning.  Even with other measures of fine-tuning in the
literature -- such as $\delhs$ which includes the effects of large
logarithms from renormalization that occur in models defined at a high
scale \cite{delhs,kn,papucci} or the traditional measure $\delbg$
\cite{eenz,bg} which most readily incorporates correlations between
weak scale parameters -- light higgsinos are a must. 
This is often hidden because proponents of these measures often emphasize
the contribution of top squark masses to fine-tuning, 
and the role of higgsinos is obscured. 
It has been noted, however, that correlations among model parameters
(such as $A_0\simeq -1.6m_0$, $m_{1/2}=(0.15-0.25)m_0$, {\it etc.})
can lead to low $\delew$ along with $\delbg \simeq \delew$
\cite{raman,bbmp} even for TeV scale top-squarks well beyond the LHC
reach: see also Ref.~\cite{casas}. Since the $\mu$-parameter occurs
directly in (\ref{eq:mz}) and renormalizes relatively little between the
high scale and the weak scale, models with large $|\mu|$ are necessarily
fine-tuned in any theory where the super-potential parameters and the
soft-SUSY breaking parameters have independent origins. For this reason,
we view light higgsinos as an essential feature of SUSY models with low
fine-tuning~\cite{rns}.

In most high scale models where parameters are defined at a scale $Q\sim
M_{\rm GUT}-M_{\rm Planck}$, the Higgs soft mass is typically driven via 
renormalization group evolution to large (TeV-scale) negative values
at the weak scale.
The value of $\mu$ is adjusted (tuned) to gain the measured value of $M_Z$ 
via the electroweak symmetry breaking constraint (\ref{eq:mz}).
As a result, in most models with relatively heavy sparticles $\delew$ 
is typically large~\cite{bbmp}.
To obtain a low values of $\delew$, however, we need $m_{H_u}^2$ 
to be driven through zero to a value close to $-M_Z^2$ at the weak scale.
Then, for moderate to large values of $\tan\beta$, we see from
(\ref{eq:mz}) that $\mu^2$ will also be close to $M_Z^2$.  A number of
mechanisms have been suggested to ensure this.  Perhaps the best known
example occurs in the hyperbolic branch/focus point (HB/FP)
region \cite{ccn,fp} within mSUGRA or its extension to include non-universal
third generation matter scalar masses and $A$-parameters \cite{sanford}.
Non-universality of soft-SUSY-breaking parameters in the Higgs sector
also allows for low values of $\mu^2$~\cite{nuhm1,nuhm2}. For instance,
allowing independent choices for GUT scale Higgs mass squared parameters
guarantees that one can adjust the high scale value of $m_{H_u}^2$ so
that it runs down to a value not far from $-M_Z^2$ at the weak
scale. Indeed, this is the idea underlying the radiatively-driven
natural SUSY (RNS) framework (described in Sec.~\ref{sec:model}) that we
adopt for our analysis. Low values of $\mu^2$ can also be obtained by
allowing non-universality in the gaugino sector. Models with high values
of the GUT scale wino mass parameter~\cite{hm2} or low values of the
gluino mass parameter~\cite{lm3} that lead to relatively low values of
$|\mu|$ have been suggested in the context of well-tempered WIMPs. 
In a recent study~\cite{martin}, Martin has suggested a model
where the SUSY breaking $F$-term also breaks the GUT symmetry via a
field that is a linear combination of a {\bf 1} and {\bf 24} dimensional
representation of $SU(5)$ leading to a non-universal pattern of GUT scale
gaugino mass parameters~\cite{wss} that results in a relatively small
magnitude of $\mu$. 

At this point one may well wonder how the little hierarchy between the
magnitude of $\mu$ (required to obtain low values of $\delew$) 
and the sparticle mass scale (typified by the gravitino mass $m_{3/2}$ in 
gravity-mediation) arises in the first place: $\mu\ll m_{3/2}$. 
The answer lies in the proposed solutions to the well-known $\mu$-problem: why 
is the magnitude of the supersymmetric $\mu$ term so much
smaller than the Planck scale? 
These solutions all involve first forbidding
the appearance of $\mu$ via some unspecified (perhaps Peccei-Quinn) symmetry, 
and then generating it via superpotential couplings to
additional visible sector fields (as in the NMSSM \cite{nmssm}) 
or via non-renormalizable interactions with hidden sector fields 
in either the K\"ahler potential
(Giudice-Masiero or GM) \cite{gm} or the superpotential (Kim-Nilles or
KN) \cite{kimnilles}.  In GM, then one expects in gravity mediation that
$\mu\sim \lambda m_{\rm hidden}^2/M_P\sim \lambda m_{3/2}$ where $\lambda$ is a
dimensionless coupling constant, while in KN then one expects $\mu\sim
\kappa f_a^2/M_P$ where $f_a$ is the Peccei-Quinn (PQ) symmetry breaking scale.
In the former case, the little hierarchy $\mu\ll m_{3/2}$ could result
from small $\lambda \sim 0.01-0.1$, while in the latter case it might 
be a reflection of a mis-match between PQ and hidden sector mass scales
$f_a\ll m_{\rm hidden}$~\cite{bbmp}.

Light higgsinos can be definitively searched for at electron-positron
colliders provided that $\sqrt{s}>2m({\rm higgsino})$.
It has been shown that at the proposed International Linear Collider
(ILC) with 80\% polarized electron beams -- in addition to discovery --
detailed measurements that unequivocally point to the higgsino origin
of the signal should be possible~\cite{rns_ilc} even in the very
challenging case where the visible energy release in higgsino decay is
as small as 10~GeV.\footnote{In models with radiatively-driven naturalness,
requiring $\delew \leq 30$ (corresponding to no worse than 3\%
electroweak fine-tuning), the mass gap between the next-to-lightest 
neutralino and the lightest SUSY particle (LSP) is always larger than 10~GeV.
Even smaller mass gaps below the GeV level can
also be explored at ILC via  a very different technique \cite{list}.} 
An ILC operating at $\sqrt{s} \simeq 600$~GeV should be able either 
to discover light higgsinos or decisively exclude the idea of natural SUSY
with better than 3\% EW fine-tuning, if gaugino mass unification is assumed.

In view of the importance of naturalness, it is critical to examine
prospects for light higgsino discovery at the LHC where the light
higgsino states $\tw_1^\pm$ and $\tz_{1,2}$ can be pair produced by
quark anti-quark collisions via the processes, $pp \to \tz_i\tz_j +X$,
$pp\to \tw_1\tz_i + X$ and $pp \to \tw_1^+\tw_1^- +X$
($i,j=1,2$). The produced charginos and neutralinos then decay via
$\tw_1 \to f\bar{f}'\tz_1$ and $\tz_2 \to f\bar{f}\tz_1$, where
$f$ and $f'$ are SM quarks and leptons, potentially resulting in
various multi-jet plus multi-lepton signatures, including the much
celebrated clean trilepton signature for SUSY. The problem, of course,
is that in natural SUSY where $|\mu|\ll |M_{1,2}|$
the higgsinos states all have a mass close to
$|\mu|$, because the mass gaps $m_{\tw_1}-m_{\tz_1}$ and
$m_{\tz_2}-m_{\tz_1}$ are proportional to $M_W^2/M_{1,2}$ and typically about 
10-20~GeV. As a result, the visible daughter jets and leptons from
charginos and neutralino decays are
very soft, and likely buried under SM backgrounds~\cite{tao,hworld}. 
The lack of a trigger for these soft events is an additional complication.

This led various groups to consider LHC prospects for discovering light
higgsinos produced in association with a hard QCD jet
\cite{monojet,Fox,tao,chinese,monous,sz,monojet-ex}, 
a photon~\cite{Fox,monous,acr,atlas-photon,cms-photon,Bramante:2014dza},
$W$ or $Z$ boson~\cite{monoWZ,monoWZ-ex} or even a Higgs boson~\cite{monoh},
where the associated particle could serve as a trigger for the
event, resulting in characteristic mono-something events with nothing
other than the very soft debris from $\tw_1$ or $\tz_2$ decays.\footnote{The 
contact interaction approximation used in many analyses of the mono-something signal 
is inapplicable for light higgsino models; see Ref.~\cite{buchmueller}.}
We are pessimistic about prospects for LHC discovery of higgsino pairs 
via detection of monojets or mono-photons. 
The problem is that although a statistically
significant signal might be possible, the signal to background ratio
is $\sim 1$\%, {\em without any discernable difference in the shapes
  of any distributions that characterize the jet or the photon}
\cite{monous,chinese,sz}.  We find it difficult to imagine that the absolute
rates for tails of QCD backgrounds will be known to better than
a percent precision required to unequivocally discover the signal.

It had been suggested~\cite{tao} that LHC detection prospects of nearly
mass degenerate higgsinos may be improved by requiring additional soft
leptons in events.  In a recent paper, Han, Kribs, Martin and
Menon~\cite{kribs} have studied this possibility in the two-lepton plus
jet channel.  Motivated by the importance of the issue of naturalness,
we felt that an independent examination of this topic was warranted.
While we broadly agree with the conclusions of Ref.~\cite{kribs} (that
the higgsino signal may indeed be observable at LHC14 with
300-1000~fb$^{-1}$), our analysis differs from that of Han {\it et al.}
in several important respects detailed in Sec.~\ref{sec:calc}. In
addition, we show that the signal may also be detectable via the
measurement of the rate asymmetry in monojet events with same flavour
dileptons and monojet events with opposite flavour dileptons. Unlike the
signal from a cut-and-count experiment, the flavour asymmetry
measurement does not suffer from the systematic uncertainty associated
with the normalization of the background.

The remainder of this paper is organized as follows. In
Sec.~\ref{sec:model} we briefly outline the RNS framework that we use
for our analysis, and set up the model line that we use for our
study. We describe our simulation of the signal and SM backgrounds along
with our analysis cuts in Sec.~\ref{sec:calc} where we also compare and
contrast our calculations with those of Han {\it et al.}~\cite{kribs}. 
Our projections for the LHC14 reach for higgsinos assuming integrated
luminosities ranging from 100-1000~fb$^{-1}$ form the subject of
Sec.~\ref{sec:reach}. We conclude in Sec.~\ref{sec:concl} with a summary
of our results and our outlook for discovering natural SUSY at the LHC.

\section{A model line with radiatively-driven naturalness}
\label{sec:model}

The RNS framework provides a setting for generating MSSM spectra with
$\delew$ in the 10-30 range. Specifically, we generate these in the
framework of the NUHM2 model~\cite{nuhm2} specified by the parameter set, 
\be 
m_0,\ m_{1/2},\ A_0,\ \tan\beta,\ \mu,\ m_A 
\ee 
consisting of the familiar mSUGRA/CMSSM parameters specified at the GUT
scale, augmented by the weak scale parameters $\mu$ and
$m_A$. The ability to specify $\mu$ and $m_A$ independently of other
parameters arises from the freedom to choose the GUT scale values of the
scalar Higgs soft-SUSY breaking parameters independently of the $m_0$,
which is why the model is referred to as the non-universal Higgs mass
model with two additional parameters.  If we choose $A_0 \sim
-(1.5-2)m_0$ then radiative corrections in (\ref{eq:mz}) from top squark
loop contributions $\Sigma_u^u(\tst_{1,2})$ are suppressed even
for $m_{\tst_1}=1-2$~TeV and $m_{\tst_2}\sim (2-4)m_{\tst_1}$. The large
magnitude of $A_0$ at the same time lifts $m_h$ to its measured value
because the top-squarks are highly mixed. Then, if we adopt $|\mu| \sim
100-300$~GeV, we find $\delew$ in the 10-30 range as desired~\cite{ltr}. 
The RNS spectrum is characterized by \cite{rns,rnslhc},
\bi
\item  light higgsino states $\tw_1^\pm$ and $\tz_{1,2}$ in the
  100-300~GeV range (the lighter the better) 
  with a mass gap between the heavier higgsinos
 and $\tz_1$ of about $10-30$~GeV,
\item well-mixed third generation squarks with TeV scale masses,
\item $m_{\tg} \leq 4-5$~TeV so that gluino loop corrections do not
  uplift the  top squark masses.  
\ei 
First/second generation sfermion masses can be chosen to be in the
5-30~TeV range without jeopardizing naturalness provided one of several
degeneracy patterns within first/second generation sfermion multiplets
is respected~\cite{maren}.  Although not required by naturalness, we
adopt this choice because it ameliorates the SUSY flavour problem by
decoupling the new physics \cite{flav}, and also addresses the proton
decay \cite{proton} and gravitino \cite{gravitino} problems (in models
where gravitinos get a mass comparable to that of these scalars).

Within the RNS framework, the mass gaps between $\tw_1/\tz_2$ and the
higgsino-like LSP $\tz_1$ decrease with increasing $m_{1/2}$. The smallest 
value of $m_{\tz_2}-m_{\tz_1}$ consistent with $\delew <
30$ is about 10~GeV~\cite{rns_ilc}. Since LHC signals for higgsino detection
become most challenging for small values of the mass gap, we adopt a
NUHM2 model line (where we vary that parameter $\mu$ which fixes the mass scale for
the higgsino-like states)
\be
  m_0=5~{\rm TeV},\ m_{1/2}=1~{\rm TeV},\ \tan\beta=15,\ A_0=-1.6m_0,\ 
  m_A=1~{\rm TeV}\;,
\label{eq:mline}
\ee
which yields $m_{\tz_2}-m_{\tz_1}$ close to  10~GeV for
our initial analysis. (For completeness, we note that the lighter chargino mass
tracks the $\tz_2$ mass and that   
$|m_{\tz_2}-m_{\tw_1}|< 2$~GeV for 100~GeV $< \mu <$~250~GeV which, as
we will see below, covers the higgsino mass range accessible at even the
high luminosity LHC.)
For the calculation of sparticle mass spectra and sparticle decay
patterns we use the ISAJET 7.84 program~\cite{isajet} where we fix
$m_t=173.2$~GeV.  Squarks and gluinos are very heavy ($m_{\tg}\sim
2.4$ TeV and $m_{\tq}\sim 5$ TeV) and so beyond the range of LHC14,
except perhaps for ultra high integrated luminosities.

\begin{figure}[tb]
\centering
\includegraphics[height=10cm]{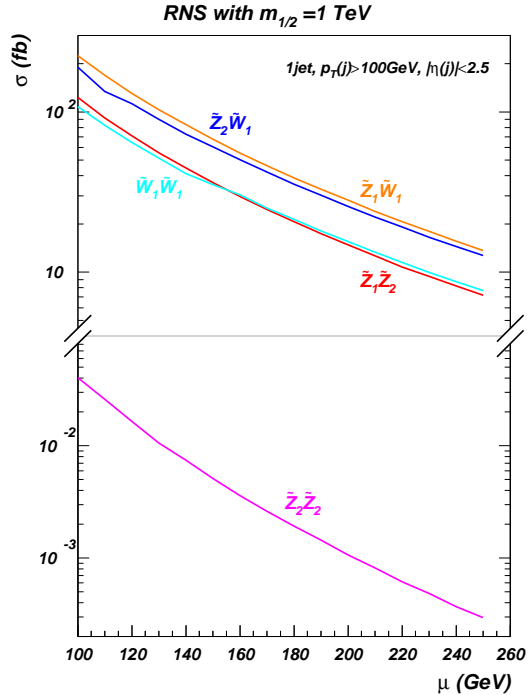}
\caption{The production cross-sections for higgsino-like chargino/neutralino
    pair plus one central jet with $p_T(j) > 100$~GeV as function of $\mu$
    for the RNS model introduced in (\ref{eq:mline}) of the text. Notice
    the break in the vertical scale.} 
\label{fig:csec1j}
\end{figure}

In Fig.~\ref{fig:csec1j}, we show cross sections along this model line
at a $pp$ collider with $\sqrt{s}=14$~TeV for the pair production of
higgsino states in association with exactly one jet with $p_T>100$~GeV jet  
and 
$|\eta_j| < 2.5$ from QCD
radiation. The details of our calculation are
described in Sec.~\ref{sec:sim}. We show results for $pp \to
\tw_1^+\tw_1^- j$, \ $\tw_1\tz_{1,2}j$ and $pp \to \tz_1\tz_2j,
\ \tz_2\tz_2j$ processes, for which soft leptons from subsequent decays
$\tw_1 \to \ell\nu\tz_1$ and $\tz_2\to \ell^+\ell^-\tz_1$ could serve
to reduce the background. We see that the cross-sections for the
production of {\em unlike} higgsino-like states (which occurs via
essentially unsuppressed electroweak couplings) in association with a
jet range from ${\cal O}(10-100)$~fb for $\mu = 100-250$~GeV. In contrast,
the cross section for $\tz_2\tz_2 j$ production (and also for
$\tz_1\tz_1 j$ production, not shown because it gives monojet events
without any additional leptons) is strongly suppressed. This is
because for $|\mu| \ll |M_{1,2}|$, the lighter neutralinos are
dominantly the higgsinos $(\th_u\pm \th_d)/\sqrt{2}$, causing the
$Z\tz_i\tz_i$ coupling to be strongly suppressed~\cite{wss}.

Since squarks and sleptons are very heavy, the $\tw_1/\tz_2$ decay
amplitudes are dominated by virtual $W/Z$ exchange so that the
leptonic branching fractions are essentially governed by $W$ or $Z$
boson couplings to quarks and leptons: 11\% (per lepton family) for
charginos, and 4\% for neutralinos for the entire range of $\mu$ shown in
the figure.\footnote{The reader should keep in mind that because
  $m_{\tz_2}-m_{\tz_1}=10$~GeV, $\tz_2 \to b\bar{b}\tz_1$ decays  are
 essentially forbidden, while decays to charm and tau fermions are 
kinematically suppressed,
  causing the leptonic branching fraction to be somewhat enhanced
  relative to that of the on-shell $Z$ boson.} The $\tw_1\tz_2j$,
$\tz_1\tz_2j$ processes, followed by the leptonic decays of $\tz_2$,
yield hard monojet plus soft dilepton plus $\eslt$ events with a cross
section $\sim (10-100)$~fb $\times 8\%$, corresponding to 
$\sim 100-1000$ events with this topology (before lepton
acceptance cuts) per 100~fb$^{-1}$ in the upcoming runs of LHC14. 
These dileptons necessarily have opposite-sign (OS) 
but same-flavour (SF) and their invariant mass is bounded from above by the
mass gap $m_{\tz_2}-m_{\tz_1} \simeq 10$~GeV along this
model-line. $\tw_1^+\tw_1^- j$ production can also lead to dilepton plus
hard monojet events at comparable rates. In this case, however, we
expect $e^+e^-$, $\mu^+\mu^-$ as well as $e^+\mu^-$ and $\mu^+e^-$
pairs at roughly the same rate. Though the dilepton mass is not
kinematically bounded as for dileptons from $\tz_2$ decays, we
nevertheless expect relatively low dilepton masses simply because the
leptons are expected to be soft. While the signal occurs at an
observable rate, there are numerous sources of SM backgrounds under
which this signal can hide. These include, $(Z/\gamma^* \to
\tau\bar{\tau})+j$, $t\bar{t}$, $W^+ W^- j$ discussed in
Ref.~\cite{kribs}, together with backgrounds from
$(Z\to\nu\bar{\nu})+(Z/\gamma^* \to \ell\bar{\ell}/\tau\bar{\tau})+j$,
$(W\to\ell/\tau+\nu)+(Z/\gamma^* \to \ell\bar{\ell}/\tau\bar{\tau})+j$, 
$(Z\to \nu\bar{\nu})+b\bar{b}+jets$ 
(where leptons from the decays of $b$ are accidentally isolated) 
and single top backgrounds from inclusive 
$tW$ and $tq'$ production. Note that backgrounds from $(Z/\gamma^* \to\ell\bar{\ell})+j$ 
events can be efficiently removed by a $\eslt$ cut.

Before closing this section, we note that $\tw_1\tz_1$ production can
give rise to single lepton plus monojet events. We expect that these
will be obscured by enormous $Wj$ and possibly top backgrounds (see also
Ref.~\cite{tao}) and do not consider these any further. $\tz_2\tw_1 j$
production potentially yields soft trilepton plus hard monojet
events. After folding in leptonic branching fractions, the total signal
rate, before acceptance cuts for leptons, is just 20-200 events per
100~fb$^{-1}$. However, since the signal leptons tend to be very soft,
we expect that the signal is severely rate limited and likely
unobservable except perhaps for the smallest values of $\mu$ at the high
luminosity LHC. For these reasons, the dilepton plus monojet signal
will be our focus in the remainder of this paper.

\section{Calculational Details}
\label{sec:calc}

\subsection{Simulation}
\label{sec:sim}

For the simulation of the signal, we use Madgraph 5~\cite{madgraph} to
generate $pp\to\tw_1^+\tw_1^-$, $\tz_{1,2}\tz_{1,2}$ and
$\tw_1^\pm\tz_{1,2}$ plus one-parton processes (exclusive) and plus
two-partons (inclusive) where, to increase efficiency, we require the
hardest final state parton to have $p_T({\rm parton})>80$~GeV; the final
cross section is then the sum of 1-jet exclusive and 2-jet inclusive
processes.  To avoid double-counting, we used the MLM scheme~\cite{mangano} for
jet-parton matching. The events are then passed to
Pythia v6.4~\cite{pythia} for showering, hadronization and underlying
event. We use CTEQ6L1 parton distribution
functions \cite{cteq} for our calculations. For the computation of the 
backgrounds we also use a similar procedure, except that for
backgrounds with an extra jet in the final state, {\it e.g.} $\tau\bar{\tau}j$, we 
only generate events with exactly one parton jet, again with $p_T>80$~GeV. 

These Madgraph/Pythia events are then passed to the Isajet toy detector
simulation~\cite{isajet} with calorimeter cell size
$\Delta\eta\times\Delta\phi=0.05\times 0.05$ and $-5<\eta<5$. The HCAL
(hadronic calorimetry) energy resolution is taken to be
$80\%/\sqrt{E}\oplus 3\%$ for $|\eta|<2.6$ and FCAL (forward calorimetry) is
$100\%/\sqrt{E}\oplus 5\%$ for $|\eta|>2.6$, where $\oplus$ denotes combination
in quadrature. The ECAL (electromagnetic calorimetry) energy resolution
is assumed to be $3\%/\sqrt{E}\oplus 0.5\%$. We use the cone-type Isajet
jet-finding algorithm~\cite{isajet} to group the hadronic final states
into jets. Jets and isolated leptons are defined as follows:
\bi
\item Jets are hadronic clusters with $|\eta| < 3.0$,
   $R\equiv\sqrt{\Delta\eta^2+\Delta\phi^2}\leq0.4$ and $E_T>40$~GeV.
\item Electrons and muons are considered isolated if they have $|\eta|
  < 2.5$, $p_T(l)>5$~GeV with visible activity within a cone of
  $\Delta R<0.2$ about the lepton direction, $\Sigma E_T^{cells}<
  {\rm min}(|p_T(\ell)|/10, 5 \; {\rm GeV})$.
\item  We identify hadronic clusters as
  $b$-jets if they contain a $B$ hadron with $E_T(B)>15$~GeV, $\eta(B)<3$ and
  $\Delta R(B,jet)<0.5$. We assume a tagging efficiency of 60\% and
  light quark and gluon jets can be mis-tagged
  as a $b$-jet with a probability 1/150 for $E_{T} \leq$ 100~GeV,
  1/50 for $E_{T} \geq$ 250~GeV, with a linear interpolation
  for intermediate $E_{T}$ values.
\ei

\subsection{Backgrounds}

We have evaluated backgrounds to the monojet plus (soft) dilepton plus
$\eslt$ signal from various SM origins:
\begin{enumerate}
\item $(Z/\gamma^* \to\tau\bar{\tau})+j \to \ell^+\ell'^-j+\eslt$, whose
  squared matrix element we evaluate as a $2\to 3$ process, and the
  taus are decayed in Pythia using the TAUOLA code~\cite{tauola};
\item $t\bar{t}$, where leptons come from decays of W and $b$ quarks;
\item $W^+W^-j \to \ell^+\ell'^-j+\eslt$;
\item $(Z\to \nu\bar{\nu})+(Z/\gamma^* \to \ell^+\ell^-)+j \to \ell^+\ell^-j+\eslt$;
\item single-top processes $bg \to tW$ and $bq \to tq'$, where the leptons come from 
  the $W$ and $b$ daughters of top decay;
\item  $(Z\to \nu\bar{\nu})+b\bar{b}+jets$, 
  where both leptons comes from the decays of the $b$ quarks.
\item $(W\to \ell/\tau+\nu)+(Z/\gamma^* \to \ell\bar{\ell}/\tau\bar{\tau})+j \to \ell \ell \ell' j+\eslt$.
\end{enumerate}

The first three of these backgrounds have also been considered in the
analysis of Ref.~\cite{kribs}. We were led to examine backgrounds from
single top production because it has a substantial cross-section:
nearly 50\% of top pair production \cite{singletop}. We will see that
it makes a sub-dominant (but non-negligible) contribution to the
background after various cuts 
to enhance the signal are implemented.  

The results of our calculations of the various
backgrounds are shown in Table~\ref{tab:back}. For reference, we also
include the corresponding signal cross section for the case with
$\mu=150$~GeV along the model line introduced in (\ref{eq:mline}).
The first row shows the total cross-section that we obtain with the
requirement of an 80~GeV parton jet in all but the $t\bar{t}$ and $tW$
production processes in the table. Also, the $t\bar{t}$ production
cross section is normalized to its NNLO+NNLL value of 953.6~pb
\cite{mitov} which corresponds to scaling the tree-level leading order
cross-section that we obtain by a factor $k=1.72$. 
Since the other backgrounds are smaller than $t\bar{t}$ and many have
smaller NLO corrections ${\cal O}(20\%)$, we did not include any
$k$-factors for them.  Although the analysis in Ref. \cite{Cullen}
suggests that NLO corrections to the signal can be large yielding
$k\sim 2.3$, which is significantly larger than $k\sim 1.4$ for
higgsino production without a jet~\cite{prospino}, we take a
conservative approach and use the LO cross section for the signal in
this study.
\begin{table}
\begin{center}
\begin{tabular}{|p{3.5cm}|r|r|r|r|r|r|r|r|}
\hline 
 & $\tau\bar{\tau}j$ & $t\bar{t}$ & $WWj$ & $Z\gamma^* j$ & $W\gamma^* j$ &
$tW$ & $tq$ & Signal \\ 
\hline 
\hline 
before cuts & 20660 & 953600 & 460.53 & 5.59 & 171.83 & 62330 & 68080 & 398.7 \\ 
1-event level & 0.011 & 0.0954 & & 0.000016 & & 0.0312 & 0.068 & \\  
\hline
$N_b=0$ & 20395 & 386450 & 455.0 & 5.54 & 169.7 & 37270 & 41550 & 393.2 \\ 
$N_j=1$ & 8020.4 & 33470 & 290.9 & 4.24 & 100.2 & 6945 & 7752 & 236.8 \\ 
$p_T(j_1)>100$~GeV, $|\eta(j_1)|<2.5$ & 2961 & 15590 & 162.7 & 1.95 & 58.8 & 2351 & 3646 & 133.7 \\ 
$\eslt > 100$~GeV & 935.0 & 12950 & 82.28 & 1.54 & 19.16 & 1700 & 2257 & 120.6 \\ 
$N_\ell\geq 2$ & 171.3 & 880.8 & 52.26 & 0.56 & 15.0 & 72.79 & 19.75 &
 2.20 \\
\hline 
$m^2_{\tau\tau}<0$ & 26.68 & 354.6 & 22.64 & 0.39 & 6.29 & 30.07 & 9.60 & 1.34 \\ 
$OS/SF$ & 12.80 & 145.9 & 11.33 & 0.39 & 4.03 & 10.94 & 3.61 & 1.19 \\ 
$m_{ll}<10$~GeV & 1.35 & 5.24 & 0.27 & 0.19 & 0.51 & 0.34 & 0.34 & 1.15 \\ 
\hline
\end{tabular}
\caption{Cut flow for the dominant backgrounds and for the RNS model-line point 
with $\mu =150$~GeV. The first row shows the cross section after the 80~GeV 
cut on the parton jet. The $WWj$, $W\gamma^* j$ and signal cross sections
are evaluated by combining several channels, making it impossible to 
assign a 1-event level.  
All cross sections are in fb. 
\label{tab:back}}
\end{center}
\end{table}

To start with, $t\bar{t}$ is by far the dominant SM process. Requiring
a $b$-veto, together with exactly one 100~GeV jet, reduces it by two
orders of magnitude, but it is still a factor 5 larger than the single
top processes, and an order of magnitude larger than 
$\tau\bar{\tau}j$ processes. Requiring $\eslt>100$~GeV together with 2
detected leptons ($e$ or $\mu$) leaves $t\bar{t}$ as the
dominant background to monojet + $\eslt$ events with a pair of soft
leptons.  This differs qualitatively from the results in Table I of
Ref.~\cite{kribs} where $\tau\bar{\tau}j$ yields the
dominant background.  While we find the $t\bar{t}$ in
Table~\ref{tab:back} to be 881~fb, the corresponding number is 28~fb
in Ref.~\cite{kribs}. We have traced this difference to several factors.
\bi
\item We use more conservative $b$-tagging criteria so that 
  the $b$-jet veto reduces the cross section by just 2.5 rather than 5 used 
  by Han {\it et al.}
\item In contrast to Ref.~\cite{kribs} which has included only $t\to
  b\ell\nu$ ($\ell=e,\mu$) decays of both tops, we include leptons
  from {\em all decay modes} of the $t$-quarks. We have checked that
  this increases the cross-section by a factor $\sim 2.25$ (of which
  just $\sim 1/3$ comes from $W\to \tau\to\ell$ decays). Somewhat surprisingly, 
  including semileptonic decays from the $b$-daughters almost doubles the rate!
  This is presumably partly due to combinatorial factors, and partly due to the 
  fact that it is easier for soft leptons to be isolated. 
\item Ref.~\cite{kribs} uses the tree-level cross-section which differs
  by a factor of $k=1.72$ mentioned earlier from the NNLO+NNLL
  cross-section that we use.
\ei

These considerations imply that we would obtain a cross-section larger
by a factor $\sim 2\times 2.25\times 1.72\simeq 8$ than that in
Ref.~\cite{kribs}, still not enough to to account for the discrepancy
between the two calculations. To better understand this, we attempted
to reproduce the numbers in Table~1 of Ref.~\cite{kribs} using the cuts
and simulation described therein. We were able to reproduce these
results for the $(W\to \ell\nu)+(W\to\ell'\nu)+j$, but are too large by about 40\% for
$\tau\bar{\tau}j$ and about a factor 3
 for the $t\bar{t}\to \ell\nu \ell'\nu b\bar{b}$ process. 
We have checked that labelling hadronic clusters with $E_T>30$~GeV as jets, as in
Ref.~\cite{kribs}, reduces the $t\bar{t}$ background by factor of 1.7. 
Extending the jet rapidity acceptance to $|\eta|<4.5$ increases the reduction to 1.9, 
accounting for the bulk of the difference of the $t\bar{t}$ background for 
that in Ref.~\cite{kribs}.

For the $WWj$ process -- 
in addition to both $W$s decaying to $\ell$ that was considered in Ref.~\cite{kribs} -- 
we also included channels where one or both $W$ bosons decay to tau:
$W\to\tau\nu\to\ell\nu\nu\nu$.  Although here one is penalized by the
leptonic branching fraction of taus, $BR(\tau\to\ell\nu\nu)\simeq
0.35$, the resulting leptons are typically soft and thus contribute
to our range of interest. 
As a result of inclusion of $W\to \tau\to \ell$ decay chains, the $WWj$ cross
section after the $N_\ell\geq 2$ cut is increased by about 20\%.

Since $\tw_1\tz_2j$ production gives up to three leptons, we
did not veto events with a third lepton to maximize the signal.  Hence, we also
need to include the 3-lepton background, $W\gamma^* j$.  Here we
include gauge bosons ($W$ or $\gamma^*$) decaying directly to $\ell$ or via 
$\tau \to \ell\nu\nu$ chain. 
We see that $W\gamma^* j$ remains about a third of $WWj$ all the way to the
$N_\ell\geq 2$ cut stage.

We mention that the $(Z\to\nu\bar{\nu})+b\bar{b}+jets$ processes -- 
though they have cross-sections $\sim 1-2$~pb range after the $\eslt$ cut -- 
make a sub-fb contribution once we require $N_{\ell}\geq 2$. 
For this reason, we have
not included these processes in the table, but have retained them in
our computation of the reach in the following section.  

Notice also that the single top contributions ignored in Ref.~\cite{kribs} are
comparable to those from $\tau\bar{\tau}j$ and $WWj$ production. The bottom line of
this discussion is that the total background that we find for dilepton
+ monojet + $\eslt$ events is six times larger than the value found 
in Ref.~\cite{kribs}. Moreover, as already mentioned, the bulk of our
background comes from $t\bar{t}$ production in contrast to
$\tau\bar{\tau}j$ production as in Han {\it et al}.

\subsection{Enhancing the Signal: Additional  Analysis Cuts}

At this stage, the signal for our benchmark point with $\mu=150$~GeV
is just 2.2~fb, with a $S/B$ ratio $\sim 0.2$\% and a statistical
significance $S/\sqrt{B} \lesssim 0.6$ even for an integrated luminosity of
1~ab$^{-1}$. Clearly, other analysis cuts are needed for the
observability of the signal. With this in mind, we see that the signal
receives contributions from $\tw_1\tw_1j$, $\tz_1\tz_2j$ and
$\tz_2\tw_1j$ processes shown in Fig.~\ref{fig:csec1j}. For the
$\mu=150$~GeV RNS point, these are 0.42~fb, 0.61~fb and 1.17~fb,
respectively.\footnote{We have checked that despite the small
mass gap the leptons pass the
identification requirements with an efficiency of $\sim 25$\%. }

The bulk of the $\tau\bar{\tau}j$ background arises from $Zj$ events
and can be efficiently removed using the kinematic procedure
\cite{hinchliffe} that allows for the reconstruction of the $Z$-peak
in di-tau events where the $Z$ has a large transverse momentum,
so that the two taus are not back-to-back in the transverse
plane. Also, since the taus are ultra-relativistic, the daughter
lepton and the associated neutrinos are all boosted in the direction
of the parent $\tau$ momentum.  In the approximation that the lepton
and the neutrinos from the decay of each tau are all exactly collimated
in the tau
direction, we can write the momentum carried off by the two neutrinos
from the decay $\tau_1\to\ell_1\nu\nu$ of the first tau as
$\xi_1 \vec{p}(\ell_1)$, and likewise for the second tau.
Momentum conservation in the transverse plane requires,
$$-\vec{p}_T(j)=(1+\xi_1)\vec{p}_T(\ell_1) + (1+\xi_2)\vec{p}_T(\ell_2).$$
Since this is
really two independent equations (recall we require $p_T(j)>100$~GeV), 
it is possible to use the measured values of the jet and
lepton momenta to solve these to obtain $\xi_1$ and $\xi_2$,
event-by-event. It is simple to check that in the approximation of
collinear tau decay, the squared mass of the di-tau system is given
by
\be
m^2_{\tau\tau}=(1+\xi_1)(1+\xi_2)m_{\ell\ell}^2 .
\label{eq:taumsq}
\ee
For dilepton plus jet events from $Z$-decay to taus, we expect
$\xi_i>0$ and $m^2_{\tau\tau}$ to peak at $M_Z^2$. Moreover, for these
events, the missing energy vector will usually point in between the
two lepton momentum vectors in the transverse plane.  In contrast, for
backgrounds where $\eslt$ arises from neutrinos from decays of heavy
SM particles ($t$, $W$, $Z$), the lepton and $\eslt$ directions are
uncorrelated and the ${\eslt}$-vector may point well away, or even
backwards, from one of the leptons so that one (or both) $\xi_i
<0$. This is also the case for the signal where $\eslt$ mainly arises
from the undetected neutralinos. In these cases, it is
entirely possible that the di-tau squared mass as given by
(\ref{eq:taumsq}) is negative.\footnote{Han {\em et al.} handle
  negative values of $\xi_i$ very differently. For events with $\xi_i <
  0$, they scale the corresponding neutrino energy by $-\xi_i$ so the
  neutrino four-vector is always time-like and $m_{\tau\tau}^2 >
  0$. From our vantage point, $m_{\tau\tau}^2$ in (\ref{eq:taumsq}) is
  the di-tau squared mass only for $Z\to \tau\bar{\tau}$ decays to fast
  taus. For all other events it is merely a kinematic quantity that
  facilitates the separation of this background from the signal. We do
  not, therefore, flip the sign of the neutrino energy when
  $\xi_i<0$. We expect that our variable is a better discriminator of
  the signal from $Z\to \tau\bar{\tau}$ events because the procedure
  in Ref.~\cite{kribs} accidentally places some negative $\xi_i$ events
  into the $Z$-peak. }

\begin{figure}[tb]
\centering
\includegraphics[height=0.4\textheight]{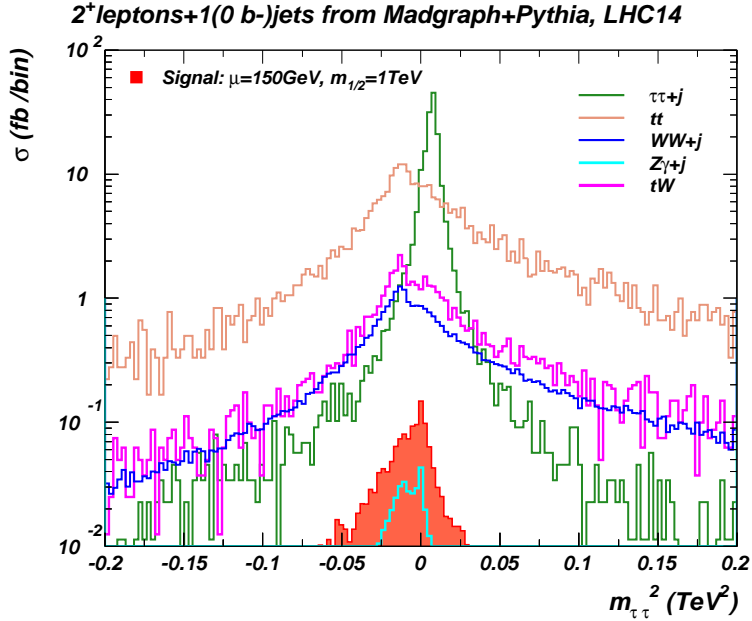}
\caption{The distribution of $m^2_{\tau\tau}$ as defined in
  (\ref{eq:taumsq}) for the main SM backgrounds to the monojet plus
  soft-dilepton plus $\eslt$ signal is shown as open histograms. The
shaded histogram denotes the corresponding distribution from the signal
for the point on the model-line (\ref{eq:mline}) with $\mu=150$~GeV.
\label{fig:mtausq}}
\end{figure}

The result of our computation of $m^2_{\tau\tau}$ for dominant SM
backgrounds as well as for the RNS model-line case with $\mu=150$~GeV
is illustrated in Fig.~\ref{fig:mtausq}.
The most striking feature is the relatively sharp peak at
$m^2_{\tau\tau}=M_Z^2=0.0083$~TeV$^2$ from $\tau\bar{\tau}+j$
events. The other backgrounds, as well as the signal, show a much
broader distribution extending to negative values of $m^2_{\tau\tau}$
for reasons just discussed. Clearly, selecting events with
\bi
\item $m^2_{\tau\tau} < 0$ 
\ei
largely eliminates the next-to-largest $\tau\bar{\tau} j$ background 
(this was the largest background in Ref.~\cite{kribs}) and
reduces other backgrounds by more than half while retaining about 60\%
of the signal.

As noted earlier, only about 20\% of the monojet plus soft dilepton
signal arises from $\tw_1\tw_1 j$ production. This falls to just 7\%
after the $m^2_{\tau\tau}< 0$ cut just discussed. The majority of the
signal thus arises from dileptons produced via leptonic decays of
$\tz_2$ in $\tz_1\tz_2j$ and $\tw_1\tz_2j$ events. Since lepton
flavour is necessarily conserved in these decays, the signal
dominantly contains same flavour (SF), opposite sign (OS) 
dileptons. In contrast, the
background leptons for all but the tiny $Z\gamma^*j$ background in
Table~\ref{tab:back} come from the decay of two different particles,
and hence are as likely to contain SF as opposite flavour (OF) leptons. 
This then leads us to require 
\bi
\item OS/SF events 
\ei 
in the signal. We
see from the penultimate row of the Table that  this halves the background with
only modest loss of signal. 

Before closing this discussion, we remark that background leptons
originating in the decays of $t$, $Z$ or $W$ will tend to be relatively
hard, so that the typical dilepton mass in these events would be $\sim
M_W$.  The invariant mass of the dilepton pair in signal events is
bounded by $m_{\tz_2}-m_{\tz_1}$. With this in mind, we show the signal
as well as background cross sections requiring, in addition,
\bi
\item $m_{\ell\ell} < m_{\ell\ell}^{\rm cut}$, \ei where
  $m_{\ell\ell}^{\rm cut}$ is a cut parameter introduced in
  Sec.~\ref{subsec:count} for the optimization of the signal.  For
  now, we fix $m_{\ell\ell}^{\rm cut} \simeq
  m_{\tz_2}-m_{\tz_1}\simeq 10$~GeV in the last row of
  Table~\ref{tab:back}.  Notice that because leptons from an off-shell
  photon tend to be soft, the background rate from $W\gamma^*j$
  becomes twice that from $WWj$ after the $m_{\ell\ell}$ cut, even
  though the $W\gamma^*j$ cross-section is just a third of the $WWj$
  cross-section prior to this cut. We see that for the sample model
  point, we have a signal of 1.15~fb versus a background of 7.73~fb,
  and so would be observable with a 5$\sigma$ significance with about
  150~fb$^{-1}$ of integrated luminosity.

The reader may be concerned by the fact that QCD resonances from $b$-
and $c$-quarkonia may yield low mass dileptons in jetty events via their
decays. While we have made no attempt to simulate this, we do not
believe this will be a serious issue because the bulk of quarkonium
production occurs via QCD processes, and is removed by the $\eslt >
100$~GeV as well as the $N_j=1$ requirements. Moreover, the quarkonium
is very likely to be part of a jet so the daughter leptons will for the
most part fail the isolation criteria. 

We have also not considered detector-dependent backgrounds from
non-physics processes such as jets faking electrons which would cause
single-lepton plus jet processes to fake the signal. The most striking
example would be $Wjj$ production, where one of the jets is the detected
hard jet, the second jet is soft, and fakes an electron, and the second
lepton arises from the decay of the $W$. Note that since the two
leptons have independent origins, the dilepton mass would typically
be tens of GeV so that only a kinematically unlikely configuration could
fake the signal. An estimate of this detector-dependent
rate is beyond the scope of the present analysis; we only mention that
the authors of Ref.~\cite{kribs} find this to be unimportant when the jet
is from a light quark or a gluon, but find a small background
from $Wb\bar{b}$ production.

\section{LHC14 reach for higgsinos in the $\ell^+\ell^- j+\eslt$ channel}
\label{sec:reach}

\subsection{Reach via cut-and-count experiments} 
\label{subsec:count}

We have seen that, with appropriate cuts described in the previous
section, higgsino pair production in association with a hard jet leads
to an observable signal in the monojet+ $\eslt$ + OS/SF dilepton
channel at LHC14. As seen from Table~\ref{tab:back}, the $m_{\ell\ell}
< m_{\tz_2}-m_{\tz_1}\simeq 10$~GeV cut was crucial for the extraction
of the signal for the $\mu=150$~GeV test case that we had examined in
the last section. However, since the neutralino mass difference is not
known {\it a priori}, the reader may legitimately question how one
would know where to set the $m_{\ell\ell}$ cut.

\begin{figure}[tb]
\includegraphics[height=0.4\textwidth]{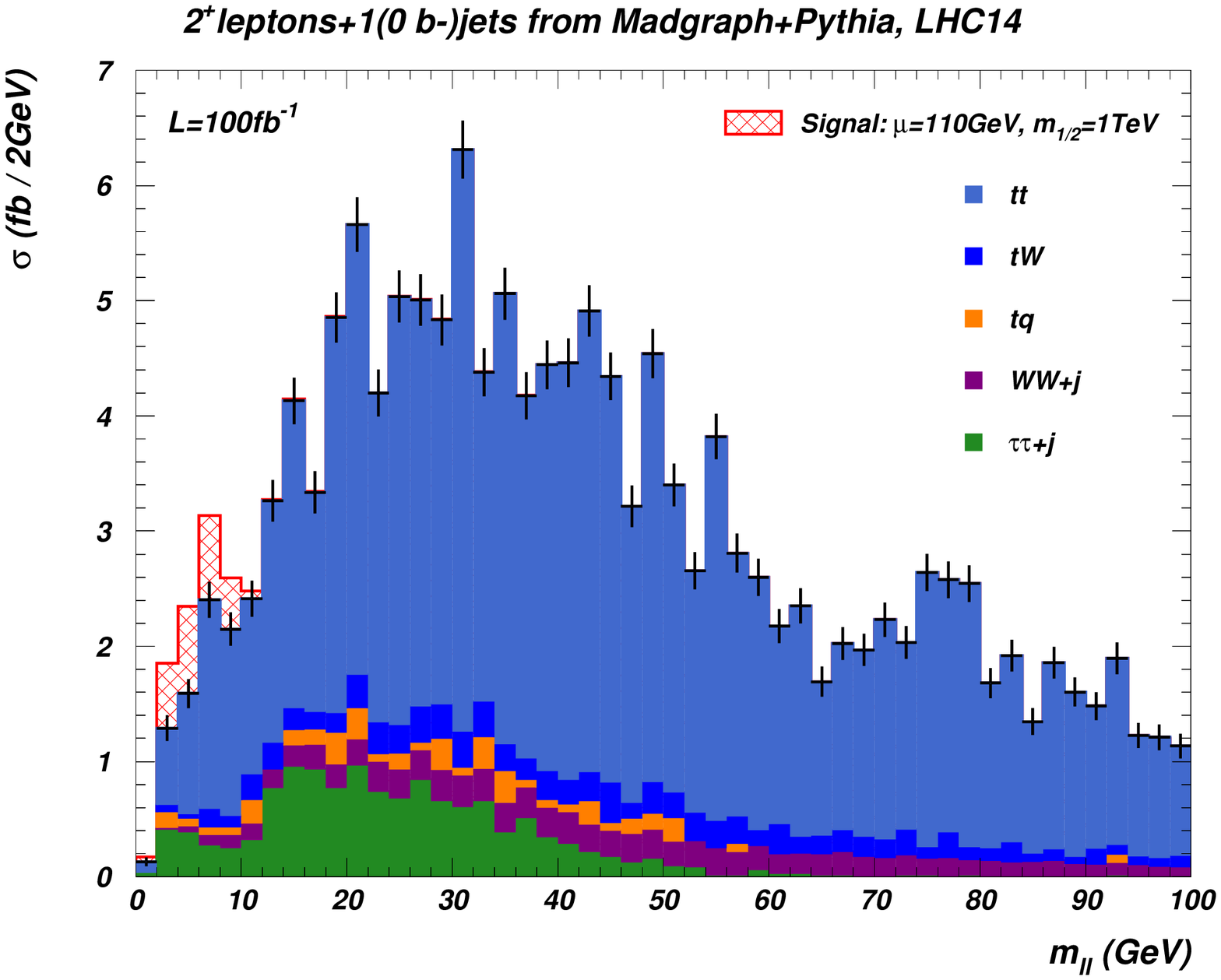}
\includegraphics[height=0.4\textwidth]{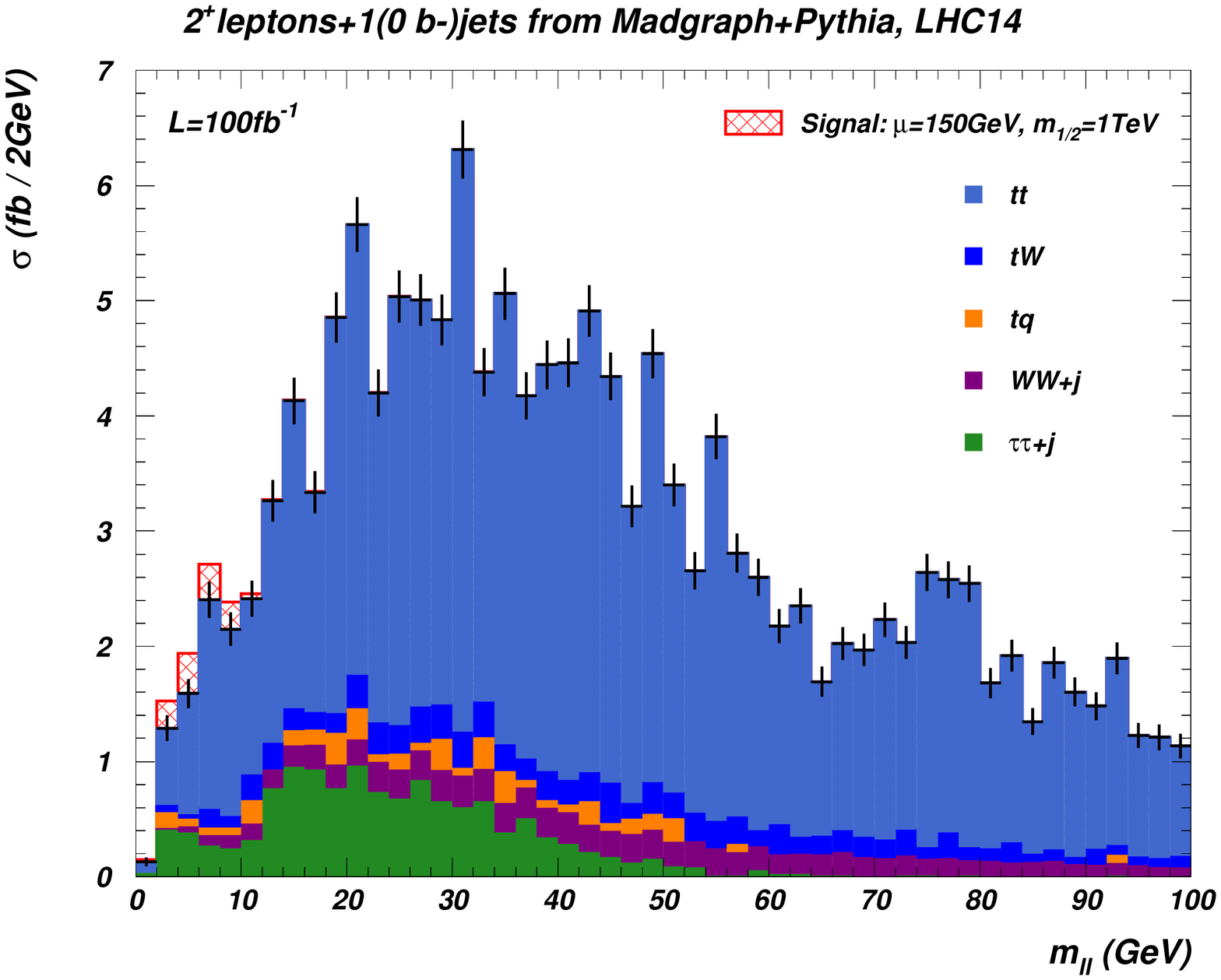}
\caption{Stacked histogram showing various SM backgrounds listed in
  Table~\ref{tab:back} of the text, together with signals for cases on
  the RNS model line (\ref{eq:mline}) for ({\it a})~$\mu=110$~GeV, and
  ({\it b})~$\mu=150$~GeV after all the cuts in this table, except
  for the requirement $0 \le m_{\ell\ell}\le m_{\ell\ell}^{\rm cut}$ in
  the last row. The background histograms (solid) are stacked in the
  same order as the legend (the small $Z\gamma^* j$ background is not
  shown for clarity) while the signal is shown as the cross-hatched
  histogram. The error bars correspond to $1\sigma$ statistical error
  on the background, assuming an integrated luminosity of
  100~fb$^{-1}$.
\label{fig:mll}}
\end{figure}

One possibility is to examine whether the invariant mass distribution
of OS/SF dileptons exhibits an obvious mass edge from the kinematic
end point of $\tz_2\to\ell\bar{\ell}\tz_1$ after all but  the
$m_{\ell\ell}$ cut in Table~\ref{tab:back}. Toward this end, in
Fig.~\ref{fig:mll} we show the stacked histogram of the $m_{\ell\ell}$
distribution from the various backgrounds in the table together with
the $1\sigma$ error bar for just the statistical error corresponding
to an integrated luminosity of 100~fb$^{-1}$. We include trilepton
events in the histogram only if the two highest $p_T$ leptons satisfy
the OS/SF requirement.  The cross-hatched histogram at low values of
$m_{\ell\ell}$ shows the contribution from the higgsino signal for the
model line in (\ref{eq:mline}) for ({\it a})~$\mu=110$~GeV, and ({\it
  b})~$\mu=150$~GeV. As anticipated, these almost cut off at
$m_{\tz_2}-m_{\tz_1}$, with the small spill-over coming mainly from
leptons from $\tw_1^+\tw_1^- j$ production.

We note the following:
\bi
\item Throughout the $m_{\ell\ell}$ range, $t\bar{t}$ production
  dominates other contributions to the background. Note that, for the
  sake of clarity, we have not included the tiny $Z\gamma^* j$ contribution
  on this histogram.
\item Due to limited resources for computing, the $t\bar{t}$ background
  simulation in the figure has been performed for an integrated
  luminosity of just $\sim$20~fb$^{-1}$.  As a result, the bin-to-bin
  fluctuations in the figure are too large by a factor of
  $\sqrt{5}$ than would be expected in a real data sample with an integrated
  luminosity of 100~fb$^{-1}$. 
  These fluctuations would be even smaller for still larger integrated
  luminosities.  Ignoring these fluctuations (which are present in the
  figure only for technical reasons) and imagining a smoother histogram,
  it is clear that it should be possible to directly extract the end
  point of the signal for the $\mu=110$~GeV case in frame {\it a},
  though a spill-over by one bin is entirely possible. For the
  $\mu=150$~GeV case in frame {\it b}, for which the signal is smaller,
  the location of the end point is less obvious.  Most certainly,
  visually locating the end point of the signal would not be possible
  for still higher values of $\mu$.
\ei

Rather than rely on such a visual determination of the end-point we,
therefore, use the following procedure to determine the selection cut.
After the cuts in all but the last row of Table~\ref{tab:back}, we
require $0\le m_{\ell\ell}\le m_{\ell\ell}^{\rm cut}$, where the parameter
$\mllcut$ is chosen to maximize the significance $S/\sqrt{B}$ of the
signal. Since our focus here is on dileptons from $\tz_2$ decays for which
$m_{\tz_2}-m_{\tz_1}$ is bounded from above, we begin by
calculating the signal and the background, choosing $\mllcut$ well
beyond the end-point for the $\tz_2\to\ell\bar{\ell}\tz_1$ decay, and
evaluating the significance of the signal. As we repeat this,
successively reducing the value of the cut parameter we see from
Fig.~\ref{fig:mll} that at first the background reduces with the
signal remaining unchanged (so that the significance increases) because
$\mllcut$ is well above the end point of the signal. However, when
$\mllcut$ goes below the kinematic end point of the signal, the signal
also starts to reduce.  Because the signal falls more sharply than the
background a maximum of the significance is obtained. It is this value
of $\mllcut$ that maximizes the significance that we use for our
projections of  the LHC reach. For the analysis shown below, we
used $\mllcut=20$~GeV to start with, and reduced it in steps of 2~GeV 
until we obtained the maximum significance. 

\begin{figure}[tb]
\centering
\includegraphics[height=0.4\textheight]{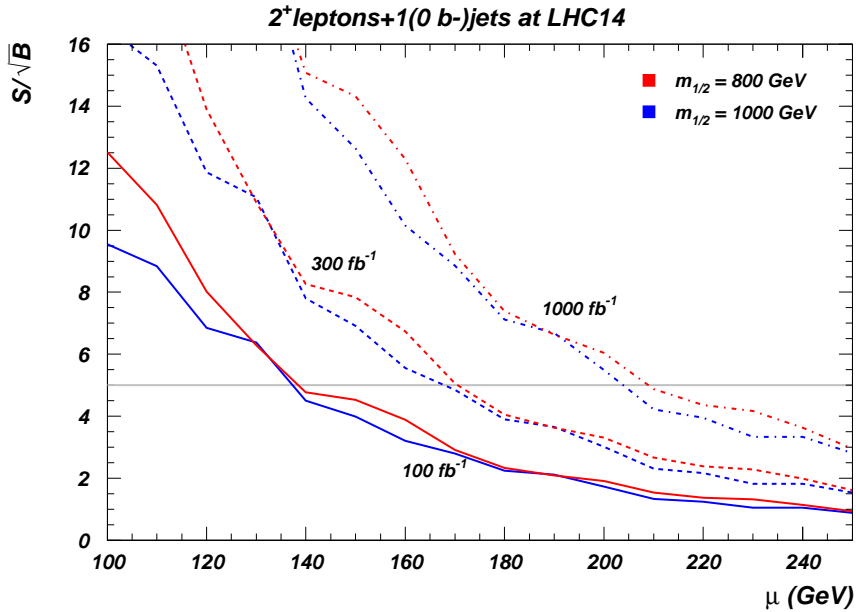}
\caption{Projections for the LHC reach via monojet plus soft dileptons
  plus $\eslt$ events from higgsinos versus $\mu$, after
  optimization using the $m_{\ell\ell}>m_{\ell\ell}^{\rm cut}$
  introduced in the text. We show results for the RNS model line of
  (\ref{eq:mline}) (lower blue curves) and also for another model line
  with $m_{1/2}=800$~GeV and the same values of other parameters
  (upper red curves), and for integrated luminosities of
  100~fb$^{-1}$, 300~fb$^{-1}$ and 1000~fb$^{-1}$. The grey horizontal
  line shows our projection for the $5\sigma$ reach neglecting any
  systematic uncertainty from the overall normalization of the
  background.
\label{fig:cutreach}}
\end{figure}
We show our projections for the LHC14 reach for the RNS model line
(\ref{eq:mline}) -- after the optimization of the $m_{\ell\ell}$ cut
just described -- {\it vs.} $\mu$ for integrated luminosities of
100~fb$^{-1}$, 300~fb$^{-1}$ and 1000~fb$^{-1}$ in
Fig.~\ref{fig:cutreach}.  LHC14 experiments are expected to accumulate
an integrated luminosity of 100~fb$^{-1}$ (300~fb$^{-1}$) in a year's
running at design luminosity (by the end of the run due to commence in
2015). We have also shown results extrapolated to 1~ab$^{-1}$ integrated
luminosity in anticipation of the high-luminosity (HL) LHC upgrade being
considered as part of the European Strategy for Physics \cite{eurstrat},
naively assuming the same detector performance in the HL
environment. For each value of integrated luminosity, we show two curves
in the figure. The lower one corresponds to $m_{1/2}=1000$~GeV as in
(\ref{eq:mline}), whereas the upper one corresponds to $m_{1/2}=800$~GeV
for which the $\tz_2-\tz_1$ mass gap ranges between 15-20~GeV (compared
to $\sim 10$~GeV for the model line in (\ref{eq:mline})) for which the
signal leptons are expected to be harder.\footnote{We have checked that
for the $m_{1/2}=800$~GeV model line $m_{\tz_2}-m_{\tw_1}$ ranges
between 4~GeV ($\mu=100$~GeV) and about 1~GeV ($\mu=250$~GeV). We note
that the precise value of the $\tw_1 -\tz_1$  mass gap is relatively
unimportant because the bulk of the signal -- after the OS/SF and
$m_{\ell\ell}$ cuts -- arises from $\tz_2 \to \ell\bar{\ell}\tz_1$
decays.} We make the following remarks:
\bi
\item We see that for both cases, experiments at LHC14 will be able to
  discover light higgsinos via monojet plus $\eslt$ events with
  identified soft dileptons in them. The reach depends on the
  integrated luminosity that is attained, and should extend out to
  $\mu\simeq 170$~GeV by the end of the next run by when an
  integrated luminosity of 300~fb$^{-1}$ is anticipated, and to
  somewhat beyond 200~GeV at the HL LHC. While this includes the least
  fine-tuned parameter space of scenarios, it does not cover all
  models where fine-tuning is limited to 3\%.
\item For the $m_{1/2}=1000$~GeV case, the optimized value of $\mllcut$ 
  always turns out to be 10~GeV for the entire range of the figure, whereas
  for the $m_{1/2}=800$~GeV case the optimized value switches from
  $\mllcut=12$~GeV for lower values of $\mu$ to $\mllcut=14$~GeV for
  larger values in the figure.  
\item Although we may expect the lepton acceptance, and hence the
  signal, to increase for the $m_{1/2}=800$~GeV case, we see that the
  LHC reach is essentially unaltered. This is because though the
  signal indeed does increase, the background which is rising for
  $m_{\ell\ell}$ values in the range of interest also increases,
  leaving the significance almost unaltered. 
\item Although our results for the reach are qualitatively similar to
  those of Han {\it et al.}, this agreement is somewhat fortuitous.
  The background that we find is several times larger, and the OS/SF
  and $m_{\ell\ell}$ cuts turn out to be crucial in our analysis.  
\item  We note that we have been rather
  conservative in the criteria that we use for $b$-jet tagging which
  determine the efficiency with which
  $b$-jets can be  vetoed.  Recall that we assume a tagging
  efficiency of 60\% for events with $b$-jets in the fiducial region. If
  $b$-jets can be more efficiently vetoed, backgrounds originating
in top quarks (recall that these form the bulk of the background in
  Table~\ref{tab:back}) will be reduced, resulting in a correspondingly
  larger reach.
\ei 

Before closing this section, we should mention that over the entire
range of $\mu$ in Fig.~\ref{fig:cutreach} where the signal has a
significance of more than $5\sigma$, the signal to background ratio
ranges from about 36\% for low values of $\mu$ down to about 5\% for
$\mu> 200$~GeV and $m_{1/2}=800$~GeV. While the distortion of the
shape of the $m_{\ell\ell}$ spectrum due to the signal is evident for
low values of $\mu$, this is {\em not the case} for the upper end of
the detectable range of $\mu$. It is, therefore, evident that
systematic errors on the {\em normalization of the background} will
have to be controlled at considerably better than the 5\% level in
order to be able to claim detection of new physics. We anticipate that
this will likely be feasible by direct measurements of the largest
backgrounds -- $t\bar{t}$ and $\tau\bar{\tau}j$ production -- in
control regions, and their extrapolation to the signal
region.\footnote{It should be possible to extrapolate the precisely
  measured $t\bar{t}$ cross sections in the inclusive 1 and 2 lepton
  channels with hard leptons to the soft dilepton channel with just
  one identified jet. Also the measurement of $Z(\to \ell^+\ell^-)+j$
  cross sections should be straightforward to translate to the
  $Z(\to\tau\bar{\tau})+j$ rates, and together with the known decay
  properties of $\tau$, to the soft-lepton plus monojet signal of
  interest.} The residual systematic using this procedure will, of
course, reduce the reach from that shown in Fig.~\ref{fig:cutreach}.

\subsection{Reach via lepton flavour asymmetry}
\label{subsec:asymm} 

We have just seen that monojet events with soft dileptons lead to a
potentially observable signal above large SM backgrounds from
$t\bar{t}$, $\tau\bar{\tau}j$, di-boson plus jet and single top production. 
It is evident though that while the signal dileptons mostly have the same
flavour $e^+e^-$ and $\mu^+\mu^-$, each of the backgrounds listed
above include an equal number of SF and OF lepton pairs. Indeed,
within the SM, the expectation of $n_{\rm SM} \equiv
N_{\rm SM}(SF)-N_{\rm SM}(OF)$ would be zero, but for the tiny contribution
from the $Z\gamma^*j$ background in Table~\ref{tab:back}. 
Since $n_{\rm sig} \agt n_{\rm SM}$, one might naively suppose that
the difference between total rates $n\equiv N(SF)-N(OF)$ would be 
a good observable to discriminate the SUSY signal from SM background. 

The dilepton flavour asymmetry,
\be
{\cal A_F}=\frac{n}{N(SF)+N(OF)}
\label{eq:asym}
\ee
has an added advantage that it is relatively insensitivity to the
overall normalization of the background, and so does not suffer from
the systematic uncertainty mentioned at the end of the previous
section. The flavour asymmetry has a statistical
  uncertainty $\delta {\cal A_F} =1/\sqrt{N(SF)+N(OF)}$, so 
that the statistical significance of the flavour asymmetry signal is
given by,
\be
\frac{{\cal A_F}}{\delta {\cal A_F}}= \frac{n}{\sqrt{N(SF)+N(OF)}}\;. 
\label{eq:signA}
\ee 
Here, $N(SF)$ and $N(OF)$ are the number of same flavour and opposite
flavour dilepton + jets +$\eslt$ events in the sample after all cuts
are applied.  Neglecting the small number of dilepton events from the
SUSY signal in the denominator, we see that the significance in
(\ref{eq:signA}) is {\it smaller} than the significance of the
corresponding cut-and-count experiment detailed in Sec.~\ref{subsec:count}
by a factor of about $\sqrt{2}$. We nevertheless feel that the flavour
asymmetry is worthy of study as it is free of the systematic from the
overall normalization of the background.

\begin{figure}[tb]
\centering
\includegraphics[height=0.4\textheight]{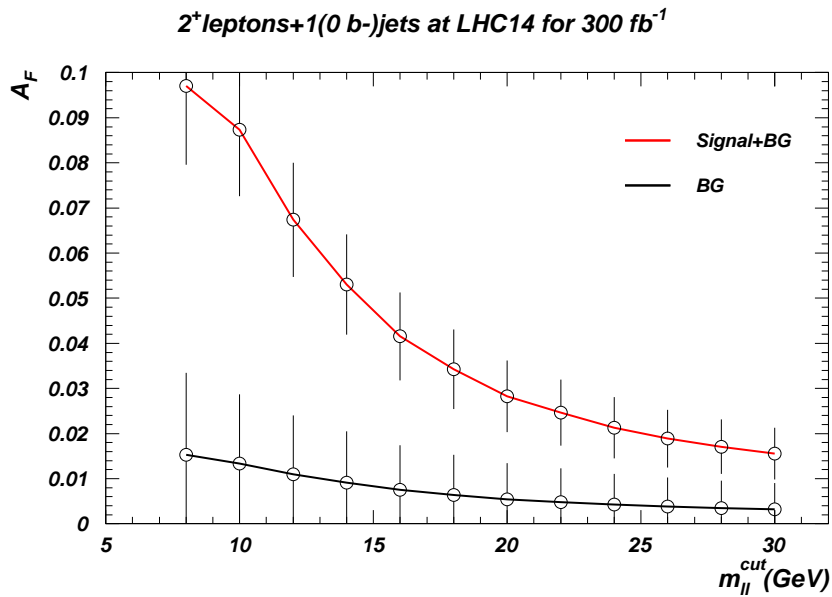}
\caption{The flavour asymmetry ${\cal A_F}$ introduced in
  (\ref{eq:asym}) as a function of the cut parameter
  $m_{\ell\ell}^{\rm cut}$ for the SM (lower curve), and for the
  higgsino signal plus background for the $\mu=150$~GeV case on the
  RNS model line (\ref{eq:mline}) (upper curve), after the cuts in
  Table~\ref{tab:back} except, of course, the OS/SF cut in the
  penultimate row, but with the requirement $0<m_{\ell\ell}<
  m_{\ell\ell}^{\rm cut}$. The error bars are for an integrated luminosity 
of 300~fb$^{-1}$ at LHC14.
\label{fig:asym}}
\end{figure}

In Fig.~\ref{fig:asym} we show the flavour asymmetry ${\cal A_F}$
after all cuts other than the OS/SF requirement in
Fig.~\ref{fig:mll}, as a function of $m_{\ell\ell}^{\rm cut}$. We show
results for the SM (lower curve), and for the SUSY signal plus
background (upper curve) for the RNS model-line (\ref{eq:mline}) case
with $\mu=150$~GeV. Also shown are statistical error bars corresponding
to an integrated luminosity of 300~fb$^{-1}$. In the SM the asymmetry
arises dominantly from $Z(\to\nu\nu)\gamma^*j$ events which have a
very small cross-section as seen in Table ~\ref{tab:back}. The rate
for these events peaks at small values of $m_{\ell\ell}$ because of
the photon pole. All other backgrounds lead to OF and SF events at equal
rates (and so no flavour asymmetry), and mostly yield dileptons with large
invariant masses, causing the asymmetry to drop with increasing value
of $m_{\ell\ell}^{\rm cut}$. We see though that the SM asymmetry is
always smaller than 1.5\% for $m_{\ell\ell}^{\rm cut}>8$~GeV. The
flavour asymmetry from the higgsino production arises via dileptons
from $\tz_2\to\tz_1\ell\bar{\ell}$ decay and is also diluted if
$m_{\ell\ell}^{\rm cut} > m_{\tz_2}-m_{\tz_1}$. Since the rate for the
signal substantially exceeds that for the $Z\gamma^* j$ background, the
asymmetry expected is significantly larger than in the SM. Indeed, we
see from Fig.~\ref{fig:asym} that with a judicious choice
of $m_{\ell\ell}^{\rm cut}$, LHC14 experiments -- with an
integrated luminosity of 300~fb$^{-1}$ -- should be able to distinguish
the flavour asymmetry produced via higgsinos from that in the SM at about
the $5\sigma$ level.

\begin{figure}[tb]
\centering
\includegraphics[height=0.4\textheight]{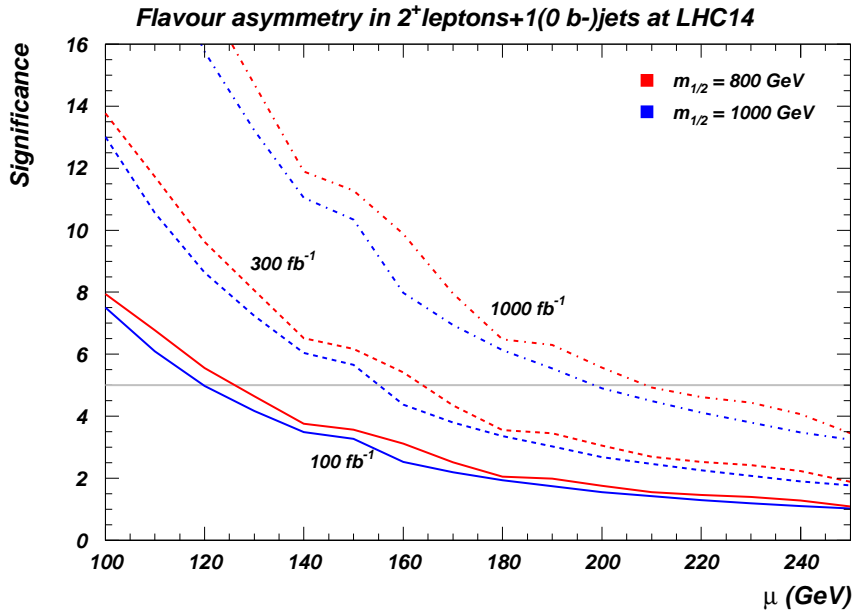}
\caption{Our projections for the LHC14 reach for higgsinos
via measurements of flavour
asymmetry 
as a function of $\mu$ after
  optimization using the $m_{\ell\ell}>m_{\ell\ell}^{\rm cut}$
  introduced in the text. We show results for the same two model lines
as in Fig.~\ref{fig:cutreach} for  integrated luminosities of 
100~fb$^{-1}$, 300~fb$^{-1}$ and 1000~fb$^{-1}$. }  
\label{fig:asymreach}
\end{figure}

To obtain the optimal value of $m_{\ell\ell}^{\rm cut}$, we follow the
procedure used in Fig.~\ref{fig:cutreach}: {\it i.e.} we choose
$m_{\ell\ell}^{\rm cut}$ that maximizes the statistical significance
${\cal A_F}/\delta {\cal A_F}$. Our projection for the LHC14 reach  via
the flavour asymmetry determination is shown in Fig.~\ref{fig:asymreach}.
We see that LHC14 experiments will be able to discover higgsinos at $5\sigma$ for
$\mu$ values up to $\sim 120$~GeV, $\sim 160$~GeV and $\sim 205$~GeV
for integrated luminosities of 100~fb$^{-1}$, 300~fb$^{-1}$ and
1000~fb$^{-1}$, respectively.  These projections are 10-15~GeV smaller
than the corresponding reach shown in Fig,~\ref{fig:cutreach} because
of the additional factor of $\sim \sqrt{2}$ in the statistical
significance ${\cal A_F}/\delta {\cal A_F}$ as discussed just below
Eq.~(\ref{eq:signA}).  We remind the reader that the result in
Fig.~\ref{fig:asymreach} will, however, not be sensitive to the
systematic uncertainty from the normalization of the background which
would, of course, reduce the corresponding reach in
Fig.~\ref{fig:cutreach}.

\section{Concluding remarks}
\label{sec:concl}

The most conservative expression of naturalness as embodied in
Eq. (\ref{eq:mz}) requires that the weak scale value of the higgsino
mass $\mu$ be not to far from the measured value of $M_Z$.
Because high scale gaugino mass parameters can be as large as $\sim
2$~TeV without jeopardizing naturalness, the existence of four light
higgsino-like electroweak -inos with compressed spectra is a robust
feature of natural SUSY.

While natural SUSY may be discovered at LHC14 via gluino pair production
\cite{rnslhc} or same-sign diboson production \cite{lhcltr,rnslhc}, the
reach in these channels (assuming gaugino mass unification) is
restricted to $m_{\tg}\alt 2$ TeV (depending on assumed integrated
luminosity).  Meanwhile, direct pair production of higgsinos can occur
with cross sections in the 10-100 fb range.  However, these are
difficult to detect because the energy release from higgsino decays is
typically very small.
In models with gaugino mass unification, the mass
gap between the second lightest and the lightest (higgsino-like)
neutralinos can be as small as 10~GeV if $\delew^{-1}$
is limited to 3\%. As a result, higgsino pair
production at the LHC (even if it could somehow be triggered on) would
be buried under large SM.

This led several groups to examine the possibility of detecting higgsino
pair production events produced in association with a central, hard jet
from QCD radiation. While there is indeed an observable event rate for
these ``monojet events'' (monojet, because the decay products of heavier
higgsinos are very soft), this strategy suffers from the fact that the
signal, which is typically just 1\% of the background, does not have any
kinematic features that help to distinguish it from $Z(\to\nu\nu)+j$
production in the SM. The
normalization of the background would, therefore, have to be known at 
considerably better than the percent level to be able to claim a
signal, a daunting prospect for tails of backgrounds of QCD origin.

Giudice {\it et al.} \cite{tao} suggested that it may be possible to
reduce the background by requiring additional soft leptons in these
monojet events.  Following this lead, Han {\it et al.} \cite{kribs}
concluded that with the additional requirement of two soft
leptons, LHC experiments should be sensitive to the higgsino signal
above SM backgrounds from $(Z\to\tau\tau) +j$ production, from
$t\bar{t}$ production and from $WWj$ production even if the neutralino
mass gap is relatively small. Prior to this study, it was
widely believed that the ILC was essential for the discovery of
light higgsinos of natural SUSY if the mass gap is small.

The importance of the LHC discovery potential for light higgsinos led us
to revisit the issue in this paper. We used the RNS framework (outlined
in Sec.~\ref{sec:model}) that we have previously developed for
phenomenological analyses of natural SUSY to guide our thinking.  For
our analysis, we re-evaluated the backgrounds listed by Han {\it et al.}
along with several other significant background sources: see
Table~\ref{tab:back}. With similar cuts, the total background that we
find is about 25 times that in Ref.~\cite{kribs}.  We have described our
attempt to track down the causes of the difference in
Sec.~\ref{sec:calc}.  In light of this order-of-magnitude larger
background, we were led to more detailed cuts, including the OS/SF and
$m_{\ell\ell}$ requirements for the signal dileptons in order to extract
the signal.  Remarkably, after a lengthy analysis, we find a higgsino
reach, as measured by the reach in $|\mu|$, qualitatively similar to
that in Ref.~\cite{kribs}. 
With an integrated luminosity of 300~fb$^{-1}$ -- targeted for 
accumulation before the shutdown for the high luminosity upgrade of the LHC --
experiments should be sensitive to higgsinos at the $5\sigma$ level if
$|\mu| < 170$~GeV, while the high luminosity upgrades should be
sensitive to $\mu$ values as high as 210~GeV: see
Fig.~\ref{fig:cutreach}.\footnote{We reiterate that, since top pair
production is by far the dominant background in Table~\ref{tab:back},
our projections for the reach are sensitive to our assumptions about how
well we can veto $b$-jets.  Here, we have conservatively assumed a
tagging efficiency of just 60\% for $b$-jets in the fiducial region, for
the upcoming LHC run, as well as for the high luminosity upgrade of the
LHC.}  It should be emphasized that the sensitivity in this figure is
evaluated using just the expected statistical error in the background
which is assumed to be measured or calculated, and does not include any
error from the uncertainty of the background. Inclusion of this
systematic uncertainty will reduce the signal significance, and hence
the reach. This could be relevant especially at the high-luminosity
upgrade where the signal to background ratio for $\mu \sim 200$~GeV is
just 5\%.
 
In addition to projecting the LHC14 reach of higgsinos from a
traditional cut-and-count
experiment, we have also evaluated the LHC reach obtained via the
measurement of the flavour asymmetry variable introduced in
Eq.~(\ref{eq:asym}). We see from Fig.~\ref{fig:asymreach} that the LHC
reach for $\mu$ via this measurement is just 10-15~GeV smaller than
that in Fig.~\ref{fig:cutreach}.
This procedure, however, has the important advantage that this reach is
insensitive to systematic uncertainties from the background
normalization.  We also see from Fig.~\ref{fig:asymreach} that, in
contrast to the reach from the counting experiment in
Fig.~\ref{fig:cutreach}, the reach via the flavour asymmetry
measurement increases with increasing neutralino mass gap.

We cannot overstress the importance of the direct search for higgsinos as
these could well be the only accessible superpartners at LHC14.
Gluinos and
first generation squarks may be in the multi-TeV range without
jeopardizing naturalness. Top squarks may also be in the 1-2~TeV
range, well beyond the LHC reach. Signals from wino production
that could lead to characteristic same-sign dilepton
events without hard jet activity~\cite{lhcltr} may well be accessible
at the high luminosity upgrade of the LHC even if gluinos
are inaccessible. But again, the accessibility of winos at even the high
luminosity LHC is not
guaranteed by naturalness considerations. 

In summary, LHC14 will be sensitive to light higgsinos of natural SUSY
models via the search for monojet events with identified OS/SF
dileptons, mostly from the decays $\tz_2\to\tz_1\ell\bar{\ell}$. These
dileptons necessarily have low invariant masses, $m_{\ell\ell}<
m_{\tz_2}-m_{\tz_1}$.  Even with an integrated luminosity of
100~fb$^{-1}$ expected from a year of LHC operation at the design
luminosity, the sensitivity of LHC14 experiments to higgsinos should
reach beyond what was explored at LEP200, and up to about what may be
observable at ILC250 \cite{rns_ilc}. The LHC reach grows with integrated
luminosity, but even with $\sim 1$ ab$^{-1}$ of integrated luminosity,
the reach in $\mu$ extends to just over 200~GeV. While the LHC can
indeed probe models of SUSY with the smallest fine-tuning, it appears
that an electron-positron linear collider with $\sqrt{s}=600$~GeV is
essential for a complete exploration of natural SUSY models with less
than 3\% electroweak fine-tuning. It is, nonetheless, exciting that
LHC14 could be a discovery machine for higgsinos of natural SUSY.

\section*{Acknowledgments}
We thank Zhenyu Han for communications about details of the 
calculations in Ref.~\cite{kribs} and for comments on the text.
This work was supported in part by the US Department of Energy.

%

%
\end{document}